\pdfoutput=1
\documentclass[11pt]{article}

\PassOptionsToPackage{hyphens}{url}

\usepackage[preprint]{acl}

\usepackage{times}
\usepackage{latexsym}
\usepackage[T1]{fontenc}
\usepackage[utf8]{inputenc}
\usepackage{microtype}
\usepackage{inconsolata}
\usepackage{graphicx}
\usepackage{amsmath}     

\usepackage[scaled=0.92]{helvet} 
\usepackage{underscore}            

\usepackage{booktabs}    
\usepackage{colortbl}    
\usepackage{xcolor}
\usepackage{verbatim}    
\usepackage{xspace}      
\usepackage{listings}    

\definecolor{ourrow}{HTML}{E8F0FE}  

\newcommand{\system}{SkillSeek\xspace}
\usepackage{url}
\makeatletter
\g@addto@macro\UrlBreaks{\do\-}  
\DeclareUrlCommand{\tool@inner}{\urlstyle{tt}}
\DeclareRobustCommand{\tool}{\tool@inner}
\makeatother

\title{\system: Revisiting Agent Skill Retrieval at Marketplace Scale\thanks{Accepted at AACL-IJCNLP 2026. This is the authors' preprint version.}}

\author{%
  Guanqun Yang\textsuperscript{1} \quad
  Wenlong Zhang\textsuperscript{1} \quad
  Tian Shi\textsuperscript{2} \quad
  Ping Wang\textsuperscript{1} \\
  \textsuperscript{1}Stevens Institute of Technology, Hoboken, NJ, USA \quad
  \textsuperscript{2}Independent Researcher \\
  \texttt{guanqun.yang@outlook.com} \quad \texttt{researchtianshi@gmail.com} \\
  \texttt{\{wzhang71, pwang44\}@stevens.edu}}

\begin{document}
\maketitle

\begin{abstract}
Anthropic's Agent Skills package reusable procedural know-how for an LLM agent into \texttt{SKILL.md} directories, and open-source aggregations have grown past 230{,}000 skills, making selection rather than authoring the bottleneck.
The standing answer in the literature outsources selection to the agent itself: an LLM-mediated retrieval loop that rewrites queries and refines candidates inside the agent's decision loop, paying LLM tokens on every task.
We present \system, an open-source two-stage skill retriever built from the standard IR recipe (a BGE-base bi-encoder feeding a small cross-encoder, exposed over MCP).
Across a $4 \times 11$ grid of pool, backbone, and method on the 89-task SkillsBench benchmark, \system\ reaches observed parity with \citet{Liu2026HowWellAgentic}'s LLM-mediated loop at essentially no extra cost: plain \texttt{bm25} alone records a pass rate at or above their refined loop on three of four settings, and a small cross-encoder covers the remaining difference on the fourth.
A first-stage recall ceiling explains the pattern, and total per-trial spend drops from \$51.30 to \$27.54 (within fifty cents of the no-skill baseline).
Under the SkillsBench tasks and OpenHands harness we tested, this positions the standard IR recipe as a strong default for agent-skill retrieval, with LLM-mediated alternatives a natural fit for cases where deterministic methods fall short.
\end{abstract}

\section{Introduction}
\label{sec:intro}

\begin{figure*}
  \centering
  \includegraphics[width=\textwidth]{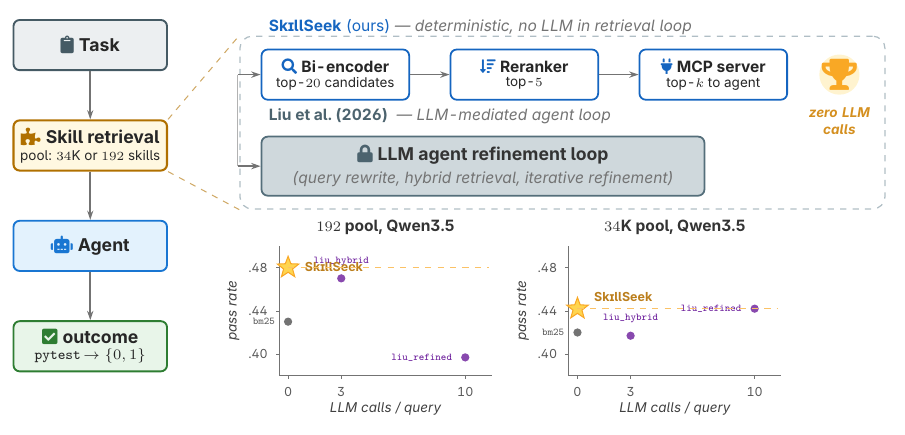}
  \caption{\textbf{A deterministic two-stage retriever reaches observed parity with an LLM-mediated retrieval loop at zero in-loop LLM cost.} Left: the task-solving loop (task $\rightarrow$ skill-retrieval module $\rightarrow$ agent $\rightarrow$ \texttt{pytest} verifier, which returns a graded reward in $[0,1]$). Right: two methods that occupy the retrieval step, \system\ (top, bi-encoder $+$ cross-encoder over MCP) and \citet{Liu2026HowWellAgentic}'s \tool{liu_refined} (bottom, LLM agent refinement loop), with cost-vs-pass-rate Pareto plots on both pools.}
  \label{fig:system}
\end{figure*}

At deployment time, an LLM can be augmented with task-specific procedural knowledge through several mechanisms: function calling,\footnote{\url{https://docs.anthropic.com/en/docs/build-with-claude/tool-use}} MCP servers,\footnote{\url{https://modelcontextprotocol.io/}} persistent project instructions like \texttt{CLAUDE.md},\footnote{\url{https://code.claude.com/docs/en/memory}} RAG,\footnote{\url{https://docs.anthropic.com/en/docs/build-with-claude/contextual-retrieval}} and sub-agents.\footnote{\url{https://docs.claude.com/en/api/agent-sdk/subagents}}
Anthropic's Agent Skills extend this toolkit with a distinctive mechanism: each skill is a \texttt{SKILL.md} directory that loads progressively (a roughly 30-token metadata entry at startup, full body when judged relevant, bundled scripts and resources on demand), leaving the agent harness to decide which skills to surface per task~\citep{Xu2026AgentSkillsLarge}.\footnote{\url{https://www.anthropic.com/engineering/equipping-agents-for-the-real-world-with-agent-skills}}
This selection problem becomes acute at scale.
Open-source aggregations of public skills have grown rapidly: 47{,}150~\citep{Li2026SkillsBenchBenchmarkingHow}, 55{,}315~\citep{Gao2026SkillReducerOptimizingLLM}, over 118{,}000~\citep{Chen2026SkVMRevisitingLanguage}, and 238{,}180 across major distribution platforms in the most recent crawl~\citep{Holzbauer2026MaliciousNotAdding}.
A natural first idea is to load the entire pool into the agent's context: modern LLMs have large enough context windows to fit the full pool.
But even at modest size (192 skills), loading every skill into the agent's context already leaves pass rate at the no-skill baseline while inflating token cost by 59\% (Appendix~\ref{sec:pilot-load-all}).
Loading a curated subset is not a free fix either: the lift from adding skills is non-monotonic~\citep{Li2026SkillsBenchBenchmarkingHow,Jiang2026SoKAgenticSkills} (SkillsBench), with one relevant skill lifting the agent by +17.8\%, two-to-three by +18.6\%, but four or more regressing to +5.9\%.
A retriever therefore has to be precise about which skills to load, not just to surface any reasonable candidate.

The standing answer in the literature is to outsource selection to the agent itself.
\citet{Liu2026HowWellAgentic} pair BM25 with a dense embedding via reciprocal-rank fusion over a 34{,}000-skill marketplace pool, and let the agent iteratively rewrite the query, explore retrieved candidates, and synthesize them into a task-specific refined skill inside its own decision loop.
The design has real strengths.
On a Terminal-Bench 2.0 tensor-parallelism case they document, the agent first retrieves two partially relevant skills (\tool{torch-tensor-parallel} and \tool{pytorch-research}).
It then composes a new skill that merges weight-sharding from the first with custom \tool{autograd.Function} patterns from the second, a synthesis that no single skill provides on its own and that a static retriever cannot produce.
When the right skills exist in the pool but no single one suffices, the LLM-mediated loop delivers something a static retriever cannot.

How often this advantage fires in practice is an empirical question.
On the SkillsBench 89-task benchmark, plain \texttt{bm25} already records a pass rate at or above \citet{Liu2026HowWellAgentic}'s refined loop on three of four (pool, backbone) settings, and \system\ (a bi-encoder $+$ small cross-encoder) covers the remaining difference on the fourth (34K pool with Qwen3.5-397B-A17B; \S\ref{sec:main}).
But accuracy parity does not extend to cost.
\citet{Liu2026HowWellAgentic}'s loop pays LLM tokens on every task, including the settings where its synthesis advantage is not needed (Table~\ref{tab:cost}): \tool{liu_hybrid} runs agentic retrieval inside the task agent's loop, inflating that loop by 44\% (\$39.43 vs.\ \$27.41 for no skills); \tool{liu_refined} adds a separate refinement agent before the task agent, totaling \$51.30 for +9.0\% over no skills.
\system\ reaches observed parity with \tool{liu_refined} at essentially no extra cost.
On the headline 34K\,/\,Qwen3.5 setting its Qwen3-Reranker-0.6B variant records the same $0.442$ pass rate as \tool{liu_refined}, while its default bge-reranker-v2-m3 variant lifts pass rate by +5.7\% over no skills against \tool{liu_refined}'s +9.0\%.
Both variants run on CPU with zero LLM tokens and leave the agent's spend within fifty cents of the no-skill floor (\$27.54 vs.\ \$27.41).

\paragraph{Contribution.}
We present \system, an open-source two-stage skill retriever (BGE-base bi-encoder feeding a small cross-encoder, exposed over MCP) implementing the standard IR recipe.
Our contribution is empirical: across a $4 \times 11$ grid of pool, backbone, and method, the deterministic recipe reaches agent pass-rate comparable to LLM-mediated retrieval, at zero in-loop LLM cost.
Under the SkillsBench tasks and OpenHands harness we tested, this positions the standard IR recipe as a strong default starting point for agent-skill retrieval, with LLM-mediated alternatives a natural fit for cases where the deterministic recipe falls short.
We open-source \system, including the scripts that regenerate every number in this paper, at \url{https://github.com/guanqun-yang/SkillSeek}.

\section{Related Work}
\label{sec:related}

\paragraph{Agent Skills.}
\citet{Li2026SkillsBenchBenchmarkingHow} contribute SkillsBench, the 89-task benchmark we evaluate on, and report that curated \texttt{SKILL.md} bundles~\citep{Xu2026AgentSkillsLarge} add +16.2\% on average across 7 model+harness configurations while bundles of 4 or more skills regress to +5.9\%.
\citet{Han2026SWESkillsBenchAgentSkills} find a similar per-skill variance: 39 of 49 public SWE skills yield zero pass-rate change; only 7 produce meaningful gains on domain-specific tasks.

\paragraph{Skill Selection and Retrieval.}
\citet{Liu2026HowWellAgentic} run the closest analogue: agentic-hybrid retrieval over a 34{,}000-skill marketplace pool, pairing BM25 with Qwen3-Embedding-4B and delegating second-stage filtering to the agent, with optional query-specific refinement.
They report pass-rate degradation from curated to marketplace-scale conditions (Claude Opus 4.6 drops from 51.2\% with curated skills to 40.1\% when retrieving from the 34K pool).
Their query-specific refinement recovers most of this loss (Claude Opus 4.6 climbs to 48.2\%) but pays an LLM-mediated exploration pass per task.
\system provides the deterministic-retrieval point of comparison alongside \citet{Liu2026HowWellAgentic}'s LLM-mediated loop, sharing a sparse-plus-dense first stage but diverging at the reranker.

Contemporaneous with this work, \citet{Zheng2026SkillRouterSkillRoutinga} study the same routing problem on a SkillsBench-derived pool of approximately 80K skills.\footnote{Posted 23 March 2026, within the three-month window of the ACL Policies for Review and Citation at our 25 May submission date. We discuss it here as concurrent work.}
They report that hiding the skill body costs 37 to 44 points of routing accuracy, and present a 1.2\,B body-aware retrieve-and-rerank pipeline reaching 74.0\% Hit@1.
Two differences separate their setting from ours: they fine-tune their own encoder and reranker, where our recipe is training-free and off-the-shelf, and they report neither a comparison against an LLM-mediated retrieval loop nor any token or monetary cost, which is the axis our contribution turns on.
Their body-is-decisive finding and our indexing ablation (\S\ref{sec:ablation}), where appending the raw skill body lowers downstream pass rate by 3.2\%, measure different quantities: routing accuracy asks whether the gold skill is ranked first, and agent pass rate asks whether the surfaced skill helps the agent finish the task.
Our helpfulness-gap diagnostic (\S\ref{sec:analysis}) shows these two can move in opposite directions.

\paragraph{Creating and Securing the Skill Pool.}
A parallel line of work builds and secures the pool.
On the authoring side, recent work compresses skill text by 48\%/39\% (description/body)~\citep{Gao2026SkillReducerOptimizingLLM}, tunes skill bundles as bi-objective search over (pass rate, cost)~\citep{Gong2026SkillMOOMultiObjectiveOptimization}, compiles skills as code for +15.3\% task completion~\citep{Chen2026SkVMRevisitingLanguage}, and mines skills automatically with 71.1\% novelty against existing libraries~\citep{Shen2026SKILLFOUNDRYBuildingSelfEvolving}.
On the trust side, guidance-injection attacks reach 16\% to 64\% trial success and evade 94\% of scanners~\citep{Liu2026TrojansWhisperStealthy}, 121 marketplace skills point to abandoned, claimable repositories~\citep{Holzbauer2026MaliciousNotAdding}, and pre-load risk scoring reaches 0.800 F1 at six tenths of a US cent per skill~\citep{Hou2026SkillSieveHierarchicalTriage}, a natural composition with \system's retrieval (\S\ref{sec:risks}).

\section{\system}
\label{sec:method}

\system is a deterministic two-stage retriever that sits between the agent and the skill pool.
We expose it as an MCP server so that any harness that already supports MCP (Claude Code, Codex CLI, OpenHands, and others) can adopt it without source-level changes.
Figure~\ref{fig:system} depicts the deployment.

\subsection{Two-Stage Retriever}
\label{sec:method-retriever}

At lookup time the agent submits a natural-language query, and \system\ runs two stages.
The first stage scores the query against every indexed skill with the BGE-base bi-encoder (110\,M parameters, 768-dim embeddings) and returns the top-20 candidates by cosine similarity.
The second stage reranks the 20 candidate (query, skill) pairs jointly with the bge-reranker-v2-m3 cross-encoder (568\,M parameters) and returns the top-5 by cross-encoder score.

The indexed text for each skill is its \texttt{name} and \texttt{description}, appended with a four-tag Tool-REX v3 structured profile (file-type, primary operation, two secondaries); the \texttt{SKILL.md} body is not indexed and is fetched only on demand.
Names and descriptions alone often miss the file-type and operation surface forms that the cross-encoder uses, while appending the raw body adds noise; the Tool-REX profile is generated offline by an LLM following \citet{Lu2025ToolsAreUnderdocumented}, and we ablate this choice against name+description and name+description+body in Table~\ref{tab:ab-indexing}.
For example, the \tool{pdf-excel-diff} skill is indexed as the concatenation shown in Listing~\ref{lst:toolrex-example}: the four tags (\texttt{pdf}, \texttt{compare}, \texttt{excel}, \texttt{xlsx}) restore exactly the surface forms the cross-encoder needs.

\begin{lstlisting}[caption={Indexed text for the \tool{pdf-excel-diff} skill under Tool-REX v3 expansion (name, description, four-tag profile).}, label={lst:toolrex-example}, numbers=left, basicstyle=\small\ttfamily, frame=single, breaklines=true]
pdf-excel-diff
Compare a PDF document against an Excel spreadsheet, flagging cells whose values differ.
pdf, compare, excel, xlsx
\end{lstlisting}

For the analysis in \S\ref{sec:analysis} we additionally swap the cross-encoder for Qwen3-Reranker variants and commercial-API rerankers; only the stage-2 model changes, while the bi-encoder, the indexing, and the top-$k$ are held fixed.
\subsection{MCP Server Interface}
\label{sec:method-mcp}

\system exposes three tools to the agent over MCP.
\tool{skill_lookup}(query, k) returns the top-$k$ candidates as \tool{name}: \tool{description} lines; \tool{skill_load}(name) returns the full \texttt{SKILL.md} body for a named skill; \tool{skill_list}() returns every skill name and description.
A server-side \tool{X-Skill-Method} HTTP header lets the experimental driver override which retriever \tool{skill_lookup} runs without changing the agent's tool schema, keeping all per-condition runs in \S\ref{sec:experiments} directly comparable across conditions.

\subsection{Driver and Harness}
\label{sec:method-driver}

The MCP server is harness-agnostic by construction.
For the experiments in this paper we drive the OpenHands SDK directly: its in-process Python agent loop lets our driver capture per-turn events, and its native MCP support lets us inject \tool{mcp_config} when swapping retrieval backends between conditions.
A discussion against alternative harnesses (Claude Code, Codex CLI, Gemini CLI, Terminus-2) is in Appendix~\ref{sec:harness-choice}.

\section{Experiments}
\label{sec:experiments}

\subsection{Experimental Setup}
\label{sec:setup}

\paragraph{Skill Pools.}
We evaluated on two pools of contrasting size and curation.
The \emph{192-skill curated pool} is the SkillsBench corpus~\citep{Li2026SkillsBenchBenchmarkingHow}: 89 deterministically-verified tasks paired with 233 raw \texttt{SKILL.md} files that deduplicate to 192 unique skill names.\footnote{A note on the related counts that appear in this paper: 89 is the full SkillsBench task set we use throughout; 233 is the total number of \texttt{SKILL.md} files across the 89 tasks' \texttt{environment/skills/} directories (approximately 2.6 skills per task); 192 is the distinct count after de-duplicating those files by name (approximately 41 skills appear in more than one task's bundle); and \citet{Li2026SkillsBenchBenchmarkingHow}'s published Table~1 uses an 84-task subset, omitting five tasks whose verifier was not yet deterministic at their snapshot date.}
The \emph{34K marketplace pool} is the open collection from \citet{Liu2026HowWellAgentic}, harvested from public skill hubs and filtered by permissive licenses and content quality.
The first pool measures the quality ceiling of retrieval when every relevant skill is in the catalog; the second measures effectiveness when the agent must find the right skill among many irrelevant ones.

\paragraph{Agent Backbones.}

The primary backbone is Qwen3.5-397B-A17B; the secondary backbone is MiniMax-M2.7, weaker by approximately 99 CodeArena points and noisier in our trials.\footnote{\url{https://llm-stats.com/leaderboards/open-llm-leaderboard}}
Both are served through OpenRouter.
Table~\ref{tab:model-ids} in the Appendix~\ref{sec:model-identifiers} lists the canonical identifier for every model used in this paper.

\paragraph{Agent Harness.}
We used the OpenHands SDK as the agent harness: it is open-source, model-agnostic, scriptable from Python, and ships native MCP and skill-loading modules, which our experimental driver depends on for swapping retrieval backends without changing the agent's tool schema.
The full comparison against alternative harnesses is in Appendix~\ref{sec:harness-choice}.

\paragraph{Methods.}
Each setting of Table~\ref{tab:main} evaluates one retrieval method paired with a fixed agent backbone over the 89 SkillsBench tasks.
All methods serve the agent's \tool{skill_lookup} tool with the top-5 skills they select from their respective pool, except \tool{liu_refined} which serves the 1 to 3 skills its refinement step selects.

We compared four families:
\begin{itemize}
  \setlength{\itemsep}{2pt}\setlength{\parskip}{0pt}\setlength{\topsep}{2pt}\setlength{\partopsep}{0pt}
  \item \textbf{Reference}: \texttt{none} mounts no skills.
  \item \textbf{Classical IR}: \texttt{bm25} alone.
  \item \textbf{OSS retriever (ours)}: a BGE-base bi-encoder over the top-20 candidates, followed by a stage-2 reranker (568\,M-parameter bge-reranker-v2-m3 or Qwen3-Reranker-0.6B); the analysis in \S\ref{sec:analysis} additionally swaps in Qwen3-Reranker-4B/8B and commercial-API rerankers (Voyage rerank-2.5, rerank-2.5-lite).
  \item \textbf{Baseline} (\citet{Liu2026HowWellAgentic}): two variants. \tool{liu_hybrid} runs their agentic retrieval (BM25 fused with Qwen3-Embedding-4B via reciprocal-rank fusion over top-60) inside the main agent's loop; \tool{liu_refined} moves that exploration into an LLM-mediated refinement subagent that runs once per task before the main agent starts. We reimplement both inside our driver (Appendix~\ref{sec:liu-faithfulness}).
\end{itemize}
The text indexed by the OSS retriever family is the skill's name and description appended with a four-tag Tool-REX v3 structured profile (file-type, primary operation, two secondaries), generated offline by an LLM following \citet{Lu2025ToolsAreUnderdocumented}; see Appendix~\ref{sec:toolrex-iter} for the exact prompt and a comparison against earlier iterations.

\paragraph{Evaluation Protocol.}
Each trial is a single agent run capped at 30 turns and 600 seconds of wall-clock.
Every SkillsBench task ships with a deterministic pytest verifier.
The verifier is deterministic but not binary: a task's assertions are scored as a group, so a trial receives a graded reward in $[0,1]$ equal to the fraction of the task's checks that pass, and partial credit is common (for example, \tool{enterprise-information-search} scores $0.200$ and \tool{weighted-gdp-calc} scores $0.593$ on a single trial).
We report the agent pass rate: the mean graded reward over the 89 tasks, with missing trials counted as 0.\footnote{Three inherited constraints shape how absolute numbers should be read: single-trial sweep at 89 tasks per setting, OpenRouter network noise on MiniMax-M2.7, and benchmark-side curation of the 192 pool. See the Limitations section.}
Table~\ref{tab:main} reports this pass rate for every setting of the main grid; first-stage retrieval recall (R@5) is reported separately in \S\ref{sec:analysis} (Table~\ref{tab:recall}).
Full harness, MCP server, and trial-budget details are in Appendix~\ref{sec:harness}.

\subsection{Main Results}
\label{sec:main}

\begin{table*}[tbp]
  \centering
\setlength{\tabcolsep}{4pt}
\small
\begin{tabular}{l l rr rr}
\toprule
& & \multicolumn{2}{c}{192 curated skills} & \multicolumn{2}{c}{34K skill marketplace} \\
\cmidrule(lr){3-4} \cmidrule(lr){5-6}
Method & Family & Qwen3.5 & MiniMax & Qwen3.5 & MiniMax \\
\midrule
\texttt{none}                  & reference                                & $0.352$            & $0.267$            & $0.352$                          & $0.267$            \\
\midrule
\texttt{bm25}                  & classical IR                             & $0.430$            & $\mathbf{0.387}$   & $0.420$                          & $\mathbf{0.346}^{\ddagger}$ \\
\midrule
\rowcolor{ourrow}
bge-reranker-v2-m3 & ours: OSS cross-encoder           & $\mathbf{0.480}$   & $0.371$           & $0.409$                          & $\mathbf{0.346}^{\ddagger}$ \\
\rowcolor{ourrow}
Qwen3-Reranker-0.6B & ours: OSS LM-based reranker  & $0.430$            & $0.353^{\dagger}$ & $\mathbf{0.442}^{\ddagger}$      & $0.341$            \\
\midrule
\tool{liu_hybrid}           & \citet{Liu2026HowWellAgentic}, no refinement                       & $0.470$            & $0.378$           & $0.417$                          & $0.335$            \\
\tool{liu_refined}          & \citet{Liu2026HowWellAgentic}, full LLM-mediated loop              & $0.397$            & $0.344$           & $\mathbf{0.442}^{\ddagger}$      & $0.329$            \\
\bottomrule
\end{tabular}

  \caption{\textbf{Main result: agent pass rate across pool, backbone, and method.} Mean pass rate over 89 SkillsBench tasks per setting (missing trials counted as 0). Our family is tinted; \textbf{bold} marks the column winner. $^{\dagger}$\,88/89: \tool{earthquake-phase-association} timed out across most MiniMax-M2.7 conditions. $^{\ddagger}$\,Within-noise ties: 34K\,/\,MiniMax and 34K\,/\,Qwen3.5 (full numbers in \S\ref{sec:analysis}). Reranker-scaling and commercial-API methods appear only on the 34K pool with Qwen3.5 (\S\ref{sec:analysis}).}
  \label{tab:main}
\end{table*}

Table~\ref{tab:main} reports the main results: agent pass rate across the $4 \times 11$ grid of (pool, backbone) pairs and retrieval methods.
\system\ is built from the standard two-stage IR recipe (bi-encoder $+$ cross-encoder); the finding we report is that this recipe reaches observed parity with the LLM-mediated retrieval loop of \citet{Liu2026HowWellAgentic}, at zero in-loop LLM cost.

\paragraph{Deterministic Methods Reach Observed Parity With \citet{Liu2026HowWellAgentic}'s Loop.}
Plain \texttt{bm25} alone records a higher pass rate than \tool{liu_refined} on three of four settings (192/Qwen3.5: 0.430 vs.\ 0.397; 192/MiniMax: 0.387 vs.\ 0.344; 34K/MiniMax: 0.346 vs.\ 0.329); only on 34K/Qwen3.5 does \tool{liu_refined} come out ahead of \texttt{bm25} (0.442 vs.\ 0.420), and there our two-stage retriever records the same value (Qwen3-Reranker-0.6B at 0.442).
Our best cross-encoder is also at or above \tool{liu_refined} on every other setting (largest observed difference +0.083 on 192/Qwen3.5: 0.480 vs.\ 0.397); on 34K/MiniMax our bge-reranker-v2-m3 (0.346) equals \texttt{bm25}.
These are observed differences on a single trial per cell, and we do not claim any of them as a resolved advantage in either direction (see Limitations).

\paragraph{Lightweight Rerankers Win on Curated Pools; Heavier Rerankers and LLM Refinement Win on Marketplace Pools.}
Which method wins changes with the size of the pool.
On the small curated pool, the simplest methods are best: \texttt{bm25} leads on MiniMax-M2.7, and our roughly 0.5\,B-parameter cross-encoder bge-reranker-v2-m3 leads on Qwen3.5-397B-A17B.
On the large marketplace pool, the heavier methods catch up: Qwen3-Reranker-0.6B (0.442) ties \tool{liu_refined} (0.442) on Qwen3.5-397B-A17B.
The same trend shows up if we track each method across pools: going from the 192 pool to the 34K pool on Qwen3.5-397B-A17B, the two lighter methods lose ground (bge-reranker-v2-m3 drops 7.1\%, \tool{liu_hybrid} drops 5.3\%), while the two heavier methods improve (Qwen3-Reranker-0.6B rises 1.2\%, \tool{liu_refined} rises 4.5\%).

\paragraph{The Same Split Holds Inside \citet{Liu2026HowWellAgentic}'s Loop.}
The split is not a property of our retriever.
\citet{Liu2026HowWellAgentic} report two variants of their own method: \tool{liu_hybrid}, which runs hybrid retrieval inside the main agent's loop with no separate refinement step, and \tool{liu_refined}, which moves that exploration into an LLM-mediated refinement subagent that runs once per task before the main agent starts.
On the 192 pool, \tool{liu_hybrid} beats \tool{liu_refined} by +7.3\% on Qwen3.5-397B-A17B (0.470 vs.\ 0.397) and +3.4\% on MiniMax-M2.7 (0.378 vs.\ 0.344); the order reverses on the 34K Qwen3.5-397B-A17B pool, where \tool{liu_refined} narrowly beats \tool{liu_hybrid} by +2.5\% (0.442 vs.\ 0.417).
LLM-mediated refinement \emph{hurts} on a small curated pool and \emph{helps} on a large noisy one.
We note that our \tool{liu_refined} is a conservative reimplementation: on the 192-pool with Qwen3.5-397B-A17B, \citet{Liu2026HowWellAgentic} report refinement \emph{helping} by +4.1\%, whereas we measure it \emph{hurting} by -7.3\% on the same backbone-pool slice; on their Kimi-K2.5 backbone they also report refinement hurting by -6.8\%, so the sign-flip from `helps' to `hurts' is something their own data already exhibits across backbones (Appendix~\ref{sec:liu-faithfulness}).

\paragraph{Takeaway.}
On the SkillsBench grid, plain \texttt{bm25} or a deterministic two-stage retriever reaches observed parity with \citet{Liu2026HowWellAgentic}'s LLM-mediated loop, at zero in-loop LLM cost.
Lightweight methods lead on the small curated pool; heavier rerankers (and LLM refinement) only catch up at marketplace scale, a pattern that holds inside \citet{Liu2026HowWellAgentic}'s own two variants.

\subsection{Ablation}
\label{sec:ablation}
\begin{table}[tbp]
  \centering
\fontsize{8.5}{10}\selectfont
\setlength{\tabcolsep}{5pt}
\begin{tabular}{l rr}
\toprule
Indexed text                          & pass rate          & $\Delta$    \\
\midrule
name $+$ description                  & $0.338$            & $-2.2\%$    \\
\quad $+$ body[:$800$]                & $0.328$            & $-3.2\%$    \\
\quad $+$ Tool-REX v$3$ (default)     & $\mathbf{0.360}$   & $0$         \\
\bottomrule
\end{tabular}

  \caption{\textbf{Indexing-text ablation.} Agent pass rate on the 34K pool with Qwen3.5-397B-A17B under three choices for the text the bi-encoder and cross-encoder index. The default (Tool-REX v3) is strongest.}
  \label{tab:ab-indexing}
\end{table}

\begin{table}[tbp]
  \centering
\fontsize{8.5}{10}\selectfont
\setlength{\tabcolsep}{5pt}
\begin{tabular}{l rr}
\toprule
Stage-1 depth                         & pass rate          & $\Delta$    \\
\midrule
$k_{\text{init}} = 10$                & $0.338$            & $-2.1\%$    \\
$k_{\text{init}} = 20$ (default)      & $\mathbf{0.360}$   & $0$         \\
$k_{\text{init}} = 50$                & $0.315$            & $-4.5\%$    \\
$k_{\text{init}} = 100$               & $0.324$            & $-3.5\%$    \\
\bottomrule
\end{tabular}

  \caption{\textbf{Stage-1 candidate-depth ablation.} Pass rate as we vary $k_{\text{init}}$, the number of candidates BGE-base passes to the cross-encoder. The default $k_{\text{init}}=20$ is the best choice.}
  \label{tab:ab-kinit}
\end{table}

\begin{table}[!tp]
  \centering
\fontsize{8.5}{10}\selectfont
\setlength{\tabcolsep}{5pt}
\begin{tabular}{l rr}
\toprule
Top-$k$ to agent                      & pass rate          & $\Delta$    \\
\midrule
$k = 1$                               & $0.340$            & $-2.0\%$    \\
$k = 3$                               & $0.360$            & $0$         \\
$k = 5$ (default)                     & $\mathbf{0.360}$   & $0$         \\
$k = 10$                              & $0.369$            & $+0.9\%$    \\
\bottomrule
\end{tabular}

  \caption{\textbf{Agent-side top-$k$ ablation.} Pass rate as we vary the number of reranked candidates returned via \tool{skill_lookup}. Beyond $k=3$, additional candidates buy nothing.}
  \label{tab:ab-topk}
\end{table}

We ablate three design choices in \system\ on the 34K\,/\,Qwen3.5 setting.
(i) \textbf{Indexed text}: Tool-REX v3 by default, compared against plain name+description and name+description+body.
(ii) \textbf{Stage-1 candidate depth $k_{\text{init}}$}: 20 by default, swept from 10 to 100.
(iii) \textbf{Number of skills returned to the agent}: top-5 by default, swept from top-1 to top-10.
Tables~\ref{tab:ab-indexing}, \ref{tab:ab-kinit}, and \ref{tab:ab-topk} report the three sweeps.\footnote{Ablation sweeps were run at a different (higher) trial-parallelism than the main grid, which slightly depresses absolute pass rates. The default Tool-REX v3 $+$ $k_{\text{init}}=20$ $+$ top-5 configuration measures 0.360 here versus 0.409 in the main grid (Table~\ref{tab:main}). To remove this concurrency-induced offset, we report $\Delta$ from the 0.360 reference throughout this subsection.}

\paragraph{Indexed Text.}
Replacing Tool-REX v3 with a plain name+description baseline drops pass rate by 2.2\%; appending the raw \texttt{SKILL.md} body drops it by a further 3.2\%, as implementation detail in the body dilutes the discriminative name+description signal.
The specific four-tag schema also matters: earlier iterations (a freeform-tag variant; a library-name-tag variant) regressed on concrete tasks because they either omitted literal file-type surface forms or let library names dominate the cross-encoder.
See Appendix~\ref{sec:toolrex-iter} for the full failure-mode analysis.

\paragraph{Stage-1 Depth.}
A naive reading of the recall-ceiling argument predicts that deeper stage-1 should help, since 34K-pool R@5 is far from saturated.
The data say otherwise.
At $k_{\text{init}}=20$ the cross-encoder behaves best; halving costs 2.1\% and doubling or quintupling costs 3.5 to 4.5\%.
More candidates inject more pairwise comparisons that perturb the rank-1 choice rather than help.

\paragraph{Agent-Side Top-$k$.}
At the agent interface, $k=1$ costs 2.0\% relative to $k=5$, while $k=3$, $k=5$, and $k=10$ are tied within 1\%.
The agent commits to rank-1 most of the time but benefits from one or two backup candidates when the cross-encoder's top-1 is wrong; beyond $k=5$, additional candidates buy nothing.

\paragraph{Takeaway.}
(i) Indexing text dominates the design space: Tool-REX v3 beats name+description by +2.2\%; appending the raw body drops a further 3.2\% (noise leakage).
(ii) $k_{\text{init}}=20$ is the best stage-1 depth: halving costs 2.1\%, doubling or quintupling 3.5 to 4.5\%.
(iii) Top-3 captures essentially all the gain over rank-1: beyond $k=3$, additional candidates buy nothing.

\subsection{Analysis}
\label{sec:analysis}

\begin{table}[tbp]
  \centering
  \resizebox{\columnwidth}{!}{
\fontsize{8.5}{9.5}\selectfont
\setlength{\tabcolsep}{5pt}
\renewcommand{\arraystretch}{1.05}
\begin{tabular}{l cc cc}
\toprule
                                       & \multicolumn{2}{c}{$192$ pool}                  & \multicolumn{2}{c}{$34$K pool}                  \\
\cmidrule(lr){2-3} \cmidrule(lr){4-5}
Retrieval method                       & R@$5$              & $\Delta$                  & R@$5$              & $\Delta$                  \\
\midrule
\multicolumn{5}{l}{\emph{Stage 1 (bi-encoder retrieval, no cross-encoder)}}                                                                \\
\quad \texttt{bm25}                    & $0.546$            & n/a                       & $0.391$            & n/a                       \\
\quad BGE-base                & $0.546$            & $\pm0.0\%$                & $0.379$            & $-1.2\%$                  \\
\arrayrulecolor{black!30}\midrule\arrayrulecolor{black}
\multicolumn{5}{l}{\emph{Stage 2 (\,$+$ cross-encoder reranker)}}                                                                          \\
\quad $+$ bge-reranker-v2-m3  & $0.581$            & $+3.5\%$                  & $0.433$            & $+5.4\%$                  \\
\quad $+$ Tool-REX v3 indexing         & n/a                & n/a                       & $\mathbf{0.478}$   & $\mathbf{+9.9\%}$         \\
\bottomrule
\end{tabular}
}
  \caption{\textbf{First-stage recall ceiling.} R@5 before and after our cross-encoder reranker on both pools. The reranker gain is much larger at the 34K scale because stage-1 recall is far from saturated there, whereas on the 192 pool it is already at ceiling.}
  \label{tab:recall}
\end{table}

\begin{figure}[tbp]
  \centering
  \includegraphics[width=\columnwidth]{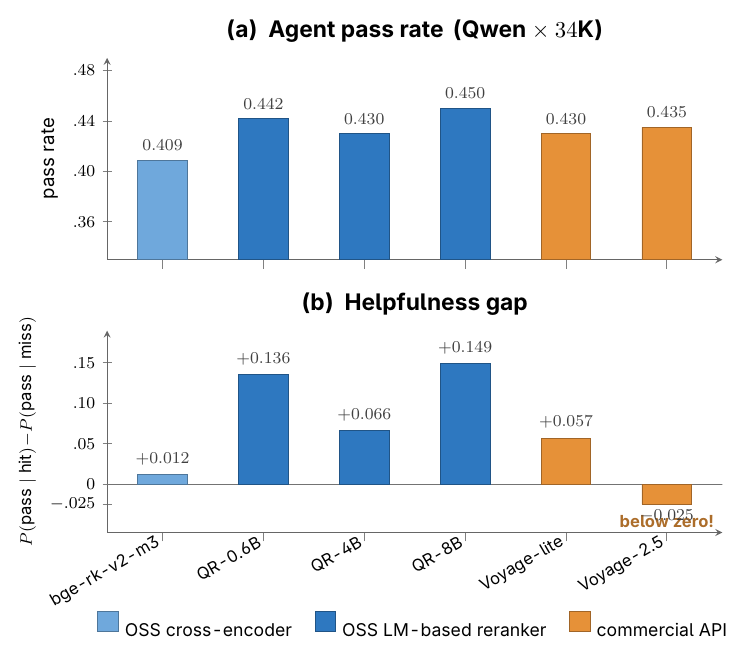}
  \caption{\textbf{Reranker scaling and OSS-vs-commercial.} 34K pool with Qwen3.5-397B-A17B. (a) Pass rate plateaus within the Qwen3-Reranker family at the 0.6B size; Qwen3-Reranker-8B narrowly beats Voyage rerank-2.5 by +1.5\%. (b) The \emph{helpfulness gap} (pass-rate conditional on the gold skill being retrieved minus pass-rate conditional on a miss) explains why: Voyage rerank-2.5 is the only reranker with a \emph{negative} helpfulness gap, meaning its retrievals do not translate into agent success.}
  \label{fig:scaling}
\end{figure}


\begin{table}[tbp]
  \centering
  \resizebox{\columnwidth}{!}{
\fontsize{8.5}{10}\selectfont
\setlength{\tabcolsep}{5pt}
\begin{tabular}{l r r r r r}
\toprule
Method                                & pass rate        & $\Delta$ vs.\ \texttt{none} & retrieval cost & agent-loop cost & total \\
\midrule
\texttt{none}                         & $0.352$          & n/a             & n/a            & \$27.41         & \$27.41   \\
\rowcolor{ourrow}
ours (BGE\,$+$\,bge-rrk-v2-m3)        & $0.409$          & $+5.7\%$        & \$0 (CPU)      & \$27.54         & \$27.54   \\
\tool{liu_hybrid}                     & $0.417$          & $+6.5\%$        & \$0            & \$39.43         & \$39.43   \\
\tool{liu_refined}                    & $\mathbf{0.442}$ & $+9.0\%$        & \$25.63        & \$25.67         & \$51.30   \\
\bottomrule
\end{tabular}
}
  \caption{\textbf{Cost breakdown.} 34K pool with Qwen3.5-397B-A17B. Retrieval cost is the pre-agent LLM spend (refinement); agent-loop cost is the main agent's own token spend on the same OpenHands SDK harness. Our deterministic retriever pays nothing on the retrieval side and leaves the agent-loop cost essentially unchanged from the no-skill baseline.}
  \label{tab:cost}
\end{table}

We now examine three questions raised by the main result: \emph{(i)} what mechanism explains why lightweight rerankers suffice on curated pools but heavier ones help on marketplace pools; \emph{(ii)} does scaling the reranker further close the gap to LLM-mediated refinement; and \emph{(iii)} how do open-source rerankers compare with commercial-API rerankers?

\paragraph{A First-Stage Recall Ceiling Explains Why Lightweight Rerankers Suffice on Curated Pools.}
We propose that the split follows from a single mechanism: \emph{cross-encoder capacity helps proportionally to how much relevant-document recall remains beyond stage-1}.
Table~\ref{tab:recall} operationalizes the claim.
On the 192 pool, \texttt{bm25} alone reaches R@5 = 0.546 and the BGE-base bi-encoder matches it; the bge-reranker-v2-m3 cross-encoder adds only +3.5\%.
Stage-1 recall is at or near ceiling: the pool was hand-built to contain the right skill, lexical metadata-matching is already sufficient, and adding capacity past bge-reranker-v2-m3 cannot lift top-1 precision.
On the 34K pool, the same comparison shows \texttt{bm25} at 0.391 and BGE-base at 0.379, far from ceiling, and the reranker adds +5.4\%, growing to +9.9\% once Tool-REX v3 further restructures the index.
This pattern is consistent with prior work.
\citet{Rosa2022DefenseCrossEncodersZeroShot} report that monoT5 capacity helps far more out-of-domain than in-domain (+0.107 nDCG@10 vs.\ +0.042 MRR over the same 60\,M to 3\,B sweep).
\citet{Li2024CanQueryExpansion} confirm the reverse direction: query-expansion makes strong cross-encoders \emph{worse} on clean test sets.
Together these predict exactly what we observe: low-capacity rerankers suffice when the right skill is already at the top of the candidate list, while high-capacity rerankers (and LLM refinement) only help when the right skill sits lower in the list and needs to be promoted.

\paragraph{Scaling Rerankers Plateaus near 0.6\,B, and Open-Source Rerankers Beat the Commercial Alternatives We Tested.}
Figure~\ref{fig:scaling}(a) reports pass rates for six rerankers on the 34K pool with Qwen3.5-397B-A17B.
Within the Qwen3-Reranker family, scaling is essentially flat: 0.6B reaches 0.442, 4B drops to 0.430 (a non-monotone dip), and 8B recovers to 0.450, a marginal +0.7\% gain over 0.6B from a 13$\times$ parameter increase.
The strongest single-reranker configuration only matches \tool{liu_refined} (0.442), consistent with the recall ceiling above bounding what any stage-2 reranker can achieve at this corpus scale.
Against the commercial alternative, Qwen3-Reranker-8B (0.450) narrowly beats Voyage rerank-2.5 (0.435) by +1.5\%, with 13 wins, 67 ties, and 9 losses on the 89-task paired comparison.

To explain why Voyage rerank-2.5 underperforms despite having the highest top-5 recall, we define a diagnostic we call the \emph{helpfulness gap}.
For a retriever $m$, partition the 89 tasks into a \emph{hit} set $G_m$ (where $m$'s top-5 contains a curated gold skill) and a \emph{miss} set $\bar{G}_m$ (where it does not); the helpfulness gap is
$$\Delta_m = \mathrm{E}_{t \in G_m}[\text{pass}_t] - \mathrm{E}_{t \in \bar{G}_m}[\text{pass}_t].$$
A positive gap means retrieving the gold helps the agent; a negative gap means it actively hurts.

Figure~\ref{fig:scaling}(b) shows Voyage rerank-2.5 with the highest top-5 recall (72.2\%) but a negative helpfulness gap ($\Delta = -0.025$): retrieving Voyage's notion of the gold skill is slightly worse for the agent than missing it.
Qwen3-Reranker-8B sits at the top ($\Delta = +0.149$); bge-reranker-v2-m3 is near zero ($\Delta = +0.012$).
The mechanism is that our agent commits to the rank-1 candidate on roughly 65\% of tasks, so what matters is the rank-1 skill being task-useful, not whether the gold is somewhere in top-5 (the AgentSkills specification leaves ranking versus browsing up to the host harness~\citep{Xu2026AgentSkillsLarge}).
Voyage rerank-2.5 was trained for general-purpose query-document relevance, so its rank-1 over-promotes semantically-similar non-gold skills; the same explains the Qwen3-Reranker-4B dip in panel (a).

\begin{table}[tbp]
  \centering
\fontsize{8}{9.5}\selectfont
\setlength{\tabcolsep}{5pt}
\begin{tabular}{l r}
\toprule
Task                                  & $\Delta$ pass rate \\
\midrule
\multicolumn{2}{l}{\emph{Top wins (rerank surfaces gold)}}    \\
\quad \tool{flood-risk-analysis}             & $+1.00$ \\
\quad \tool{pg-essay-to-audiobook}           & $+1.00$ \\
\quad \tool{gravitational-wave-detection}    & $+0.89$ \\
\quad \tool{earthquake-plate-calculation}    & $+0.88$ \\
\quad \tool{grid-dispatch-operator}          & $+0.83$ \\
\midrule
\multicolumn{2}{l}{\emph{Top losses (near-twin in pool)}}     \\
\quad \tool{gh-repo-analytics}               & $-0.25$ \\
\quad \tool{mario-coin-counting}             & $-0.33$ \\
\quad \tool{adaptive-cruise-control}         & $-0.50$ \\
\quad \tool{energy-ac-optimal-power-flow}    & $-0.65$ \\
\quad \tool{parallel-tfidf-search}           & $-0.80$ \\
\bottomrule
\end{tabular}

  \caption{\textbf{Per-task wins and losses.} Rerank vs.\ \texttt{none} on the 34K pool with MiniMax-M2.7; top-5 in each direction.}
  \label{tab:per-task-deltas}
\end{table}

\begin{table}[tbp]
  \centering
\fontsize{8.5}{10}\selectfont
\setlength{\tabcolsep}{6pt}
\begin{tabular}{l r r r r}
\toprule
Difficulty   & $n$   & \texttt{none}  & rerank   & lift     \\
\midrule
easy         & $6$   & $0.254$        & $0.306$  & $+5.2\%$ \\
medium       & $52$  & $0.324$        & $0.378$  & $+5.4\%$ \\
hard         & $26$  & $0.194$        & $0.274$  & $+8.0\%$ \\
\bottomrule
\end{tabular}

  \caption{\textbf{Per-difficulty lift.} Rerank-vs-\texttt{none} lift on the 34K pool with MiniMax-M2.7; difficulty labels come from \tool{task.toml}, with 3 of 89 tasks unlabeled.}
  \label{tab:per-difficulty}
\end{table}

\paragraph{Per-Task Wins and Losses.}
Table~\ref{tab:per-task-deltas} reports the largest per-task deltas of rerank vs.\ \texttt{none} on the 34K pool with MiniMax-M2.7.
Two patterns dominate.
The largest positive lifts (+0.89 to +1.00) occur where the gold skill's name and description match the agent's query closely; examples include \tool{flood-risk-analysis} and \tool{gravitational-wave-detection}.
The largest negative lifts (-0.50 to -0.80) occur where another pool skill is a near-twin in description and the cross-encoder picks the wrong one, as on \tool{parallel-tfidf-search} and \tool{adaptive-cruise-control}.
The losses are not coverage failures (the gold is in the pool); they are ranking failures driven by lexical near-twins that the cross-encoder lacks verifier-aware signal to disambiguate.

\paragraph{Per-Difficulty Stratification.}
Table~\ref{tab:per-difficulty} stratifies the rerank-vs-\texttt{none} lift by SkillsBench difficulty.
The lift is largest on hard tasks (+8.0\%), modest on medium tasks (+5.4\%), and similar on easy tasks (+5.2\%).
The hard-task gain is the cleanest signal: hard tasks have the lowest no-retrieval baseline (0.194) and therefore the most headroom that retrieval can convert into pass rate, consistent with the recall-ceiling argument above.

\paragraph{Cost Breakdown.}
Table~\ref{tab:cost} decomposes the total per-trial token cost into a retrieval-side component and a main-agent-loop component on the 34K pool with Qwen3.5-397B-A17B.
\system\ pays nothing on the retrieval side: the bi-encoder and cross-encoder run on CPU at approximately 1.1\,s median per \tool{skill_lookup} call (1.9\,s at p95), and the agent's own spend stays within fifty cents of the no-skill baseline (\$27.54 vs.\ \$27.41).
\tool{liu_hybrid} keeps the retrieval LLM-cost at zero but pays a 44\% surcharge on the agent loop (\$39.43 vs.\ \$27.41), because the agentic hybrid retrieval is called inside the main loop and the agent spends additional turns issuing search queries.
\tool{liu_refined} moves the agentic exploration out of the main loop into a refinement subagent that runs once per task before the main agent starts (\$25.63), which drops the main-loop spend to \$25.67 (below the no-skill baseline) but pushes the total to \$51.30 for a +9.0\% lift over no skills.

\paragraph{Takeaway.}
At marketplace scale, scaling within an open-source LM-based reranker family plateaus near 0.6\,B parameters and reaches observed parity with commercial-API rerankers and LLM-mediated refinement; stage-1 recall, not stage-2 capacity, bounds the residual difference.
The deterministic recipe records \tool{liu_refined}'s accuracy at much lower per-trial spend (\$27.54 vs.\ \$51.30), while keeping agent-side spend within fifty cents of the no-skill baseline and avoiding any in-loop LLM call.
This positions the deterministic recipe as a strong default under the SkillsBench tasks and OpenHands harness we tested, with LLM-mediated alternatives held in reserve for harder regimes.

\section{Conclusion}
\label{sec:conclusion}

We presented \system, an open-source two-stage skill retriever (BGE-base bi-encoder feeding a small cross-encoder, exposed over MCP) built from the standard IR recipe.
Across the SkillsBench 89-task benchmark, \system\ reaches observed parity with \citet{Liu2026HowWellAgentic}'s LLM-mediated loop at essentially no extra cost: plain BM25 alone suffices on three of four settings, and \system\ covers the remaining difference on the fourth, with per-trial spend dropping from \$51.30 to \$27.54.
Under the SkillsBench tasks and OpenHands harness we tested, the standard IR recipe is a strong default for agent-skill retrieval, with LLM-mediated alternatives a natural fit where it falls short.

\clearpage
\section*{Limitations}
\label{sec:limitations}

We disclose the following limitations that frame how our results should be interpreted.

\paragraph{Single-Trial Evaluation.}
Each main-grid setting is a single 89-task trial.
To estimate trial noise, we re-ran one setting (34K, MiniMax-M2.7, bge-reranker-v2-m3) three times; the standard deviation was approximately 1\%.
We therefore treat differences below 2\% as within noise, and rely on larger cross-pool effects (several percentage points) for the curated-vs-marketplace claim, not per-setting rankings.
This is why we describe our central result as observed parity with \citet{Liu2026HowWellAgentic}'s loop: we report the pass rates each method achieved, and we claim no resolved advantage in either direction.
Establishing equivalence in the stronger sense would require a pre-specified practical margin and repeated independent rollouts of the headline Qwen3.5 settings, separating variation across tasks from variation across rollouts of the same task.
We leave that replication to future work.

\paragraph{Missing Trials Are Counted as Zero.}
A trial that fails to complete, most often through an OpenRouter network error or a timeout, contributes a reward of 0 to the mean rather than being dropped.
This convention lowers every method's absolute pass rate, and it applies identically to \system, to the classical baselines, and to \citet{Liu2026HowWellAgentic}'s loop, so it does not favor any method.
It does compress the spread between methods, which makes the observed differences we report conservative in magnitude but not in sign.

\paragraph{Backbone Noise on MiniMax-M2.7.}
On MiniMax-M2.7, OpenRouter network or timeout failures hit roughly half the 89 tasks per condition, dropping absolute pass rates by 5\% to 10\% and narrowing the gap between methods.
We therefore treat Qwen3.5-397B-A17B as the headline backbone and use MiniMax-M2.7 only as supporting evidence.

\paragraph{Selection Bias in the Curated 192-Pool.}
The 192-skill SkillsBench pool was hand-curated by \citet{Li2026SkillsBenchBenchmarkingHow} so that every task has a relevant skill.
Our small-pool numbers therefore reflect both retrieval quality and curator selection.
We make no claim about the absolute 192-pool numbers; the main finding is the cross-pool pattern (lightweight rerankers suffice on the curated pool; heavier methods help on the marketplace pool).

\paragraph{First-Stage Alternatives Are Compared by Recall Only.}
Our recall-ceiling analysis (\S\ref{sec:analysis}) compares first-stage retrievers by R@5, the fraction of a task's gold skills appearing in the top five candidates.
Recall is cheap to measure because it needs no agent rollout, and it is the quantity our mechanism turns on, but it is not the quantity the paper's conclusion is stated in.
A first-stage retriever with higher R@5 need not produce a higher agent pass rate, since what matters downstream is whether the top-ranked skill is useful for the task, and our helpfulness-gap diagnostic (\S\ref{sec:analysis}) shows these two can move in opposite directions.
Running the strongest first-stage alternatives through the same cross-encoder and the same agent loop would settle this, and we leave it to future work.

\paragraph{Scope of the Claim.}
Every result here comes from the 89 SkillsBench tasks executed under one agent harness, OpenHands, with two backbones and two skill pools.
We do not claim that the standard IR recipe is a strong default for agent-skill retrieval in general, only that it is one under the tasks, harness, pools, and backbones we tested.
Terminal, web, and software-engineering agents impose different retrieval demands, and whether the observed parity survives in those settings is open.
Independently built skill-retrieval benchmarks such as SkillRet, which pairs 16{,}129 public agent skills with 4{,}392 evaluation queries~\citep{Kang2026SkillRetLargeScaleBenchmark}, are the natural targets for testing whether the recipe transfers, and we leave that to future work.



\bibliography{zotero}

\appendix

\section{Model Identifiers}
\label{sec:model-identifiers}

We refer to each model by a short, paper-readable name in the main text; Table~\ref{tab:model-ids} gives the canonical HuggingFace, OpenRouter, or Voyage identifier for reproducibility.

\begin{table}[ht]
  \centering
  \resizebox{\columnwidth}{!}{
\small
\begin{tabular}{ll l}
\toprule
Short name (paper) & Canonical identifier & Host \\
\midrule
\texttt{BGE-base}                & \texttt{BAAI/bge-base-en-v1.5}        & HuggingFace \\
\texttt{bge-reranker-v2-m3}      & \texttt{BAAI/bge-reranker-v2-m3}      & HuggingFace \\
\texttt{Qwen3-Reranker-0.6B}     & \texttt{Qwen/Qwen3-Reranker-0.6B}     & HuggingFace \\
\texttt{Qwen3-Reranker-4B}       & \texttt{Qwen/Qwen3-Reranker-4B}       & HuggingFace \\
\texttt{Qwen3-Reranker-8B}       & \texttt{Qwen/Qwen3-Reranker-8B}       & HuggingFace \\
\texttt{Qwen3-Embedding-4B}      & \texttt{Qwen/Qwen3-Embedding-4B}      & HuggingFace \\
\texttt{Voyage rerank-2.5}       & \texttt{rerank-2.5}                   & Voyage REST API \\
\texttt{Voyage rerank-2.5-lite}  & \texttt{rerank-2.5-lite}              & Voyage REST API \\
\texttt{Qwen3.5-397B-A17B}       & \texttt{qwen/qwen3.5-397b-a17b}       & OpenRouter \\
\texttt{MiniMax-M2.7}            & \texttt{minimax/minimax-m2.7}         & OpenRouter \\
\texttt{gpt-oss-120b}            & \texttt{openai/gpt-oss-120b}          & HuggingFace (vLLM) \\
\bottomrule
\end{tabular}
}
  \caption{\textbf{Canonical model identifiers.} Used in this paper across the main grid (\S\ref{sec:main}), analysis (\S\ref{sec:analysis}), and appendix.}
  \label{tab:model-ids}
\end{table}

\section{Choice of Agent Harness}
\label{sec:harness-choice}

Our retrieval conditions (\texttt{bm25}, \texttt{rerank}, \tool{liu_hybrid}, \tool{liu_refined}) are exposed as MCP servers, so the harness must (i) speak MCP natively, (ii) be scriptable from Python with open-weight backbones, and (iii) be open-source so that harness-side optimizations cannot confound retrieval comparisons.
Five candidate harnesses were available at the time of the experiments:
\begin{itemize}
  \item \textbf{Claude Code} (Anthropic) is closed-source: its skill-discovery loop is unauditable.
  \item \textbf{Codex CLI} (OpenAI) is closed-source for the same reason and tied to a single model family.
  \item \textbf{Gemini CLI} (Google) is open-source but ships only as a Node.js subprocess over stdio with no Python SDK, and is tied to the Gemini family.
  \item \textbf{Terminus-2} (used by SkillsBench evaluations and by \citet{Liu2026HowWellAgentic} for Kimi K2.5) is a deliberately-minimal benchmark baseline (approximately 1.5\,K LOC) with no native MCP and no skill module; SkillsBench added skill loading via a fork but MCP was never wired in.
  \item \textbf{OpenHands SDK} (All Hands AI) is open-source, model-agnostic, scriptable from Python, and ships native MCP and skill modules.
\end{itemize}
Of the five, only the OpenHands SDK combines an in-process Python agent loop with native MCP, which our driver depends on for per-turn event capture and direct \tool{mcp_config} injection.

\section{Harness and Implementation Details}
\label{sec:harness}

We drive the OpenHands SDK directly rather than through a wrapping CLI, with skills served over Model Context Protocol (MCP).
Each trial uses a fresh \texttt{DockerDevWorkspace} container, the agent receives the SkillsBench task instruction plus a small system-message suffix nudging it to call \tool{skill_lookup} before exploring the workspace, and the run is capped at 30 agent turns and 600 seconds of wall-clock.
The MCP server (\tool{skills_mcp.py}) exposes three tools: \tool{skill_lookup}(query, k) returns the top-$k$ candidates as \tool{name}: \tool{description} lines; \tool{skill_load}(name) returns the full \texttt{SKILL.md} body for a named skill; \tool{skill_list}() returns every skill name and description (rarely used by the agent).
A server-side \tool{X-Skill-Method} header lets the experimental driver swap which retriever \tool{skill_lookup} runs without changing the agent's tool schema, keeping all per-condition runs A/B-comparable.
For the \tool{none}, \tool{oracle}, and \tool{all} conditions of the workshop ablation we bulk-mount skills via the harness's standard \tool{AgentContext.skills} channel; for the retrieval conditions the agent calls \tool{skill_lookup} on demand.

\paragraph{Retrieval-Tool Invocation.}
In every retrieval-enabled condition we inject two instruction fragments into the agent's context.
The first is appended to the system prompt:
\begin{quote}
\itshape
IMPORTANT: A skill retrieval tool named \tool{skill_lookup} is available. Before exploring the workspace or running commands, you MUST call \tool{skill_lookup} with a short query summarizing the task to discover relevant domain skills. Then call \tool{skill_load} on the most relevant result to read its full content.
\end{quote}
The second is appended to the first user message:
\begin{quote}
\itshape
[REQUIRED FIRST STEP] Your very first tool call must be \tool{skill_lookup} with a 5 to 15 word query that summarizes this task. Then call \tool{skill_load} on the most relevant result. Only after at least one \tool{skill_load} call may you start working on the task.
\end{quote}
Empirically the system-only nudge produced roughly 40\% to 60\% \tool{skill_lookup} invocation across backbones; the dual nudge raised it to 95.5\% on every tested backbone.

\section{Baseline Reimplementation}
\label{sec:liu-faithfulness}

The main grid in \S\ref{sec:main} compares \system\ against \citet{Liu2026HowWellAgentic}'s two loops under identical conditions: same agent harness, same backbones, same task budget, same trial count.
To enable this like-for-like comparison, we reimplement \citet{Liu2026HowWellAgentic}'s \tool{liu_hybrid} and \tool{liu_refined} inside our experimental driver, drawing on their open-sourced code and prompts.
This appendix documents the reimplementation and states which way its deviations cut. Every deviation reduces what our \tool{liu_refined} can do relative to the published version, so the accuracy comparison favors \system: a fully faithful reimplementation would raise \tool{liu_refined}'s pass rates and narrow or reverse the differences we report. This is one reason we state the headline result as observed parity and claim no advantage in either direction. The cost comparison runs the other way, and in our favor: the capabilities we dropped are open-ended exploration and unlimited refined-skill output, both of which would raise \tool{liu_refined}'s token spend above the \$51.30 we measure.

\paragraph{Retrieval Server.}
Our \tool{search_server} matches Liu's behavior: BM25 fused with Qwen3-Embedding-4B via reciprocal-rank fusion at $k=60$, exposing the same \tool{/keyword}, \tool{/semantic}, \tool{/hybrid}, and \tool{/detail} endpoints.
Two minor implementation choices differ from the published source: we use the \tool{rank_bm25} Python package (no stemming) instead of SQLite FTS5, and we omit the optional content-embedding blend in the hybrid endpoint.
The agent-side discovery prompt is a verbatim port of Liu's published skill, redirected to our server port.

\paragraph{Refinement Prompt.}
The refinement prompt we use is a conservative subset of Liu's published \tool{INSTRUCTION_PROMPT}: skills are retrieved via \texttt{curl} to our server (vs.\ a pre-populated directory), exploration is capped at 10 commands (vs.\ open-ended), we drop the optional \texttt{skill-creator} meta-skill, and we cap output at 1 to 3 refined skills (vs.\ unlimited).
Each of these choices reduces what our \tool{liu_refined} can do relative to the published version.
On the 34K\,/\,Qwen3.5 cell, where Qwen3-Reranker-0.6B and \tool{liu_refined} both reach 0.442, a fully faithful reimplementation could therefore lift \tool{liu_refined} above our retriever, and readers should treat the observed parity on that cell as the optimistic reading for \system.

\paragraph{Reproduction Gap.}
On the 192-pool Qwen3.5-397B-A17B cell, \citet{Liu2026HowWellAgentic} report that query-specific refinement lifts pass rate from 0.267 to 0.308 (+4.1\%), whereas our measurement shows refinement \emph{hurting} (0.470 to 0.397, a -7.3\% delta).
This sign flip is consistent with Liu's own data: on Kimi K2.5 they also report refinement hurting (0.335 to 0.267, -6.8\%).
Our prompt's conservatism explains part of the residual gap on Qwen3.5; on the settings where we report parity, the direction of the published delta matches ours.

\section{Backbone Strength}
\label{sec:backbone-strength}

The two open-weight backbones differ in capability and stability.
Qwen3.5-397B-A17B sits at Code-Arena score 1{,}257; MiniMax-M2.7 sits at 1{,}158, an approximately 99-point gap that translates to an 8.5\% gap on our \texttt{none} baseline (0.352 vs.\ 0.267). The capability ordering is consistent across both pools and four trial sessions.
On the noise axis, the two are not comparable: MiniMax-M2.7 trials hit \tool{ConversationRunError}, \tool{TimeoutError}, or \tool{MaxIterationsReached} on 48 to 57 of the 89 tasks per condition (a method-independent error rate of roughly 50\% to 64\%), whereas Qwen3.5-397B-A17B trials see near-zero such failures.
The mechanism is straightforward: a stage-2 reranker that surfaces the most useful skill is irrelevant when the downstream agent cannot reliably execute on it. The method-level signal on MiniMax-M2.7 is therefore limited by how often the agent crashes, not by retrieval quality.
For this reason we treat Qwen3.5-397B-A17B as the headline backbone and MiniMax-M2.7 as supporting evidence, adjudicated by the lift between conditions on MiniMax-M2.7 rather than by absolute scores.

\section{Pilot Experiment on Loading the Entire Pool}
\label{sec:pilot-load-all}

\begin{table*}[h]
  \centering
\small
\begin{tabular}{l r r}
\toprule
Condition                                              & pass rate      & tokens/trial      \\
\midrule
\texttt{none} (no skills loaded)                       & 0.384          & 268\,K            \\
\texttt{all} (entire 192-skill pool loaded)            & 0.387          & 426\,K (+59\%)    \\
\texttt{oracle} (per-task curated subset, 1--3 skills) & \textbf{0.484} & 314\,K (+17\%)    \\
\bottomrule
\end{tabular}

  \caption{\textbf{Loading every skill from a small pool already fails.} On 89 SkillsBench tasks with a locally served gpt-oss-120b backbone, loading the full 192-skill pool (\texttt{all}) yields the same pass rate as loading nothing (0.387 vs.\ 0.384) despite a 59\% token surcharge, while a per-task curated subset (\texttt{oracle}) lifts pass rate to 0.484.}
  \label{tab:load-all-fails}
\end{table*}

The numbers in Table~\ref{tab:load-all-fails} come from a pilot sweep we ran prior to the main grid (\S\ref{sec:experiments}) with three conditions (no skills, the full 192-skill pool loaded into context, and the per-task curated subset of 1 to 3 skills) on the same 89 SkillsBench tasks.
The agent backbone was gpt-oss-120b served locally via vLLM on the two-RTX-6000 stack described in Appendix~\ref{sec:compute}; the harness was the OpenHands SDK with skills served over MCP, identical to the main-grid setup.
Each setting aggregates three trials per task (267 trials per condition).
Token totals are agent-side spend on gpt-oss-120b for the full trial (system prompt, task instruction, agent turns, and tool responses), averaged across trials.
We treat the pilot as supporting evidence rather than a primary result.
A stronger backbone might absorb more of the noise from loading all 192 skills, but the directional finding holds: an exhaustive load is not a viable substitute for selection.
This is consistent with the main-grid evidence that even \texttt{bm25} outperforms loading the full pool.

\section{Tool-REX Profile Iteration}
\label{sec:toolrex-iter}

Tool-REX v3 is the third iteration of the prompt used to expand each skill's index text with structured tags.
We generated all three iterations offline with a locally served gpt-oss-120b (vLLM); the index is computed once per pool and reused across all retrieval trials at zero in-loop LLM cost.
We document the two earlier failures because they map cleanly to known failure modes of LLM-generated document expansion~\citep{Li2024CanQueryExpansion}.

\paragraph{v1 (Freeform Tags): Surface-Form Drift.}
The first prompt asked the LLM for 3 to 5 freeform descriptive tags per skill.
On \tool{pdf-excel-diff}, this produced tags like \texttt{file comparison, tabular data, document inspection} but omitted the literal surface forms \tool{excel} and \tool{xlsx}.
The cross-encoder, which is sensitive to query-document surface overlap, then ranked the gold skill at position 2 or 3 instead of 1, dropping R@5 from 1.0 to 0.5 on this task and similar ones.

\paragraph{v2 (File-Type + Libraries + Operations): Library-Name Hijack.}
The second prompt added explicit file-type and library-name tags, e.g., \texttt{pdf pypdf reportlab pdfplumber}.
At marketplace scale this dominated the cross-encoder for any \tool{pdf}-shaped query: the literal \tool{pdf} skill's tag list was dominated by Python library names rather than the term \tool{pdf} itself, and library tags appeared on near-twin skills that were not the gold.
\tool{court-form-filling} dropped from R@5 $=1.0$ to 0.0 at 34K scale.

\paragraph{v3 (Final): Four Ordered Tags, No Library Names.}
The shipping prompt produces exactly four tags in fixed order: file-type, primary operation, two secondaries.
No library-name tags.
This is the version used as the indexing default in our main grid (\S\ref{sec:main}): on the 34K pool with MiniMax-M2.7 it lifts the agent pass rate by +12.8\% over a name+description baseline, with paired 24 wins, 5 losses, and 41 ties on the shared subset.

\section{Ablation Details}
\label{sec:ablation-detail}

This appendix provides per-parameter mechanism analysis for the three sweeps in \S\ref{sec:ablation}; the tables themselves (Tables~\ref{tab:ab-indexing}, \ref{tab:ab-kinit}, \ref{tab:ab-topk}) appear in the main text.

\paragraph{Indexing Text Matters; Tool-REX Is the Right Deterministic Choice.}
Replacing the Tool-REX v3 index with a plain name+description baseline drops pass rate by 2.2\%, and appending raw body text drops it by a further percentage point (-3.2\%).
The body-text loss is consistent with noise leakage: implementation detail in the body dilutes the discriminative name+description signal that the cross-encoder relies on.
The specific four-tag schema also matters: two earlier iterations of the Tool-REX prompt regressed on concrete tasks (a freeform-tag variant dropped R@5 on \tool{pdf-excel-diff} from 1.0 to 0.5 by omitting the literal surface forms \tool{excel}/\tool{xlsx}; a library-name-tag variant dropped \tool{court-form-filling} from 1.0 to 0.0 at 34K scale by letting Python library names dominate the cross-encoder).
See Appendix~\ref{sec:toolrex-iter} for the full failure-mode analysis.
Tool-REX v3 is the indexing default behind every \system\ entry in Table~\ref{tab:main}.

\paragraph{Stage-1 Depth $k_{\text{init}}=20$ Is the Best Choice; Deeper Hurts.}
A naive reading of the recall-ceiling argument would predict that on the 34K pool, where stage-1 R@5 is far from saturated, deepening stage-1 should help the cross-encoder lift relevant skills.
The data say otherwise.
The reranker behaves best at $k_{\text{init}}=20$; halving the depth costs 2.1\%, and doubling or quintupling it costs 3.5 to 4.5\%.
The shape is consistent with the rank-1-commit framing of \S\ref{sec:analysis}: when the agent loads only the rank-1 candidate, what matters is the cleanliness of the very top of the cross-encoder ranking, not raw recall depth.
More candidates in the cross-encoder's input set introduce more pairwise comparisons that can perturb the rank-1 choice.

\paragraph{The Rank-1-Commit Framing Holds, with a Few Backup Candidates.}
At the agent interface, $k=1$ costs 2.0\% relative to $k=5$, while $k=3$, $k=5$, and $k=10$ are tied within 1\%.
The agent commits to rank-1 most of the time but does benefit from one or two backup candidates when the cross-encoder's top-1 is wrong.
Beyond $k=5$, additional candidates buy essentially nothing.

\section{Helpfulness-Gap Diagnostic: Sign Interpretation}
\label{sec:helpfulness-gap}

The formal definition of the helpfulness gap $\Delta_m$ is given in \S\ref{sec:analysis}.
The sign of $\Delta_m$ has three readings.
A positive $\Delta_m$ means ``retrieving the gold helps the agent''; a near-zero $\Delta_m$ means the retrieval signal is washed out by other factors (agent noise, execution difficulty); a negative $\Delta_m$ means the retrieved gold actively confuses the agent relative to a miss.
Voyage rerank-2.5's $\Delta = -0.025$ in Figure~\ref{fig:scaling}(b) is the clearest negative case in our data.


\section{Risks}
\label{sec:risks}

Skill retrieval systems can amplify risks present in the underlying skill pool.
\system\ surfaces whichever skills the index rates as most relevant; it does not audit, sandbox, or verify them.
A malicious or low-quality skill in the pool can be surfaced and then executed inside the agent's loop; recent attacks demonstrate guidance-injection via skill bootstrap hooks~\citep{Liu2026TrojansWhisperStealthy} and abandoned-repository hijacks on public marketplaces~\citep{Holzbauer2026MaliciousNotAdding}.

\system\ inherits this exposure in full: a poisoned skill in the pool will be ranked alongside legitimate ones.
A natural risk-aware extension would compose \system's retrieval with a pre-load risk scorer~\citep{Hou2026SkillSieveHierarchicalTriage} to drop high-risk candidates before they reach the agent.

The retriever itself introduces no new attack surface beyond the underlying pool.
It runs locally, makes no external network calls beyond the LLM provider the host harness already contacts, and exposes only three read-only MCP tools (\tool{skill_lookup}, \tool{skill_load}, \tool{skill_list}); none of these write to the filesystem or modify agent state outside the standard MCP request-response channel.
The retriever also does not persist user queries, agent state, or task content across runs.

\section{Use of Existing Artifacts}
\label{sec:artifacts}

Our code is released at \url{https://github.com/guanqun-yang/SkillSeek}.

We use two existing artifacts to construct the evaluation grid in \S\ref{sec:experiments}.
The first is the SkillsBench task suite and skill pool of \citet{Li2026SkillsBenchBenchmarkingHow}, distributed under the Apache-2.0 license.\footnote{\url{https://github.com/benchflow-ai/skillsbench}}
The second is the 34{,}000-skill marketplace pool of \citet{Liu2026HowWellAgentic} (code\footnote{\url{https://github.com/UCSB-NLP-Chang/Skill-Usage}}, dataset\footnote{\url{https://huggingface.co/datasets/Shiyu-Lab/Skill-Usage}}), with no license file present at the time of access (2026-05-24).
Neither artifact ships an explicit ``intended use'' statement.
We use both as agent-skill retrieval benchmarks, consistent with the framing in the original papers.

\paragraph{Statistics.}

SkillsBench contains 89 deterministically verifiable tasks paired with 233 \texttt{SKILL.md} files (approximately 2.6 skills per task) that deduplicate to 192 unique skill names; \citet{Li2026SkillsBenchBenchmarkingHow}'s published Table~1 uses an 84-task subset that omits 5 tasks whose verifier was not yet deterministic at their snapshot date, and we evaluate on the full 89.
The Skill-Usage marketplace pool of \citet{Liu2026HowWellAgentic} contains approximately 34{,}000 skills harvested from public hubs and filtered by permissive licenses and content quality.
Both pools are English-only.
\system\ is not trained on either pool, so we use each pool as a single evaluation set with no train/dev/test split.

\paragraph{Documentation.}
SkillsBench's 89 tasks span scientific computing (e.g., \tool{flood-risk-analysis}, \tool{gravitational-wave-detection}, \tool{earthquake-phase-association}), data processing (e.g., \tool{pdf-excel-diff}, \tool{parallel-tfidf-search}), and applied software (e.g., \tool{adaptive-cruise-control}, \tool{court-form-filling}); the per-task difficulty labels we use in Table~\ref{tab:per-difficulty} come from \tool{task.toml} in each task directory, with 3 of 89 tasks unlabeled.
The Skill-Usage 34K pool covers a broader application surface drawn from public skill hubs; \citet{Liu2026HowWellAgentic} document the harvesting and filtering procedure in their appendix.

\paragraph{PII and Offensive-Content Audit.}
We manually audited a random sample of 5 SkillsBench tasks and 50 marketplace skills for personally identifiable information and for offensive content; we found no significant issues in either dimension.
The sample is small, and we make no population-level claim about either pool; we report the audit only as a sanity check on a subset we read.

\section{Computational Experiments}
\label{sec:compute}

\paragraph{Model Sizes.}
The bi-encoder (BGE-base) has 110\,M parameters and produces 768-dim embeddings; the default cross-encoder (bge-reranker-v2-m3) has 568\,M parameters.
The reranker scaling sweep of \S\ref{sec:analysis} uses Qwen3-Reranker-0.6B, Qwen3-Reranker-4B, and Qwen3-Reranker-8B, with parameter counts matching their names.
The two agent backbones in the main grid are Qwen3.5-397B-A17B (397\,B total, 17\,B active per token) and MiniMax-M2.7; both are queried through OpenRouter.
The Tool-REX v3 index expansion was generated offline by gpt-oss-120b served via vLLM.
All canonical model identifiers are listed in Table~\ref{tab:model-ids}.

\paragraph{Compute Budget and Infrastructure.}
A local vLLM stack with two NVIDIA RTX 6000 GPUs serves the two locally hosted models in this paper: gpt-oss-120b for the Tool-REX v3 index expansion, and Qwen3-Embedding-4B as the embedding component of \citet{Liu2026HowWellAgentic}'s baseline retriever.
The Tool-REX v3 index expansion required under 2 GPU-hours on this stack and is computed once per pool, then reused across all retrieval trials.
At lookup time the bi-encoder and cross-encoder of \system\ run on CPU at approximately 1.1\,s median (1.9\,s at p95) per \tool{skill_lookup} call (\S\ref{sec:analysis}); the BGE-base embedding of the 34K pool is built once on the same workstation and reused thereafter.
Agent rollouts run through OpenRouter; per-trial token cost is reported in Table~\ref{tab:cost}, and per-backbone OpenRouter spend for the full set of experiments in this paper is broken down in Table~\ref{tab:api-spend}.

For the commercial-reranker sweep in \S\ref{sec:analysis} (Figure~\ref{fig:scaling}), Voyage AI provides 200 million free tokens per account for both rerank-2.5 and rerank-2.5-lite.\footnote{\url{https://docs.voyageai.com/docs/pricing}}
Our 89-task sweep did not exhaust this one-time allowance, so the prepaid \$50 credit on the account was not drawn down.

\begin{table}[ht]
  \centering
  \resizebox{\columnwidth}{!}{
\small
\begin{tabular}{l r}
\toprule
Backbone & OpenRouter spend (USD) \\
\midrule
Qwen3.5-397B-A17B \emph{(primary)}             & 617.77 \\
MiniMax-M2.7 \emph{(secondary)}                & 149.19 \\
GLM 5 \emph{(exploratory)}                     & 45.84  \\
Kimi K2.6 \emph{(exploratory)}                 & 24.67  \\
\midrule
Total                                          & 837.47 \\
\bottomrule
\end{tabular}
}
  \caption{\textbf{OpenRouter API spend per backbone.} Totals for the four LLM backbones queried during this work. The two locally served models (gpt-oss-120b for the Tool-REX v3 index expansion and Qwen3-Embedding-4B for \citet{Liu2026HowWellAgentic}'s baseline retriever) ran on the two-GPU vLLM stack described above and are not on OpenRouter. Captured from the OpenRouter usage dashboard on 2026-05-24.}
  \label{tab:api-spend}
\end{table}

\paragraph{Package Versions and Parameter Settings.}
The BM25 baseline uses \tool{rank_bm25} with default tokenization (no stemming and no FTS5-style filtering, as documented in Appendix~\ref{sec:liu-faithfulness}).
The bi-encoder and cross-encoder are loaded through \tool{sentence-transformers} at their HuggingFace defaults; we do not fine-tune.
The retrieval interface uses the official \tool{mcp} Python SDK, and the agent loop uses the OpenHands SDK with the harness configuration documented in Appendix~\ref{sec:harness}.
The Tool-REX v3 prompt is served via vLLM with greedy decoding (temperature 0, max\_tokens capped to 256 per skill); the full prompt is reproduced in Appendix~\ref{sec:toolrex-iter}.

\section{GenAI Disclosure}
\label{sec:genai}

We maintained complete human authorship over the intellectual core of this study.
We used Claude Code to assist with typesetting tasks (such as building tables, diagrams, and plots) and LLMs to refine our wording, but no AI was used to conceptualize ideas, generate data, or perform evaluations.
We designed all algorithms, wrote the experimental code, and drafted the manuscript entirely ourselves.

\end{document}